\documentstyle[12pt]{article}
\title{
Non-adiabatic effect in electron screening in fusion reactions at
astrophysical energies\\
}
\author{K. Hagino,$^1$ N. Takigawa,$^1$ and M.Abe$^2$\\ \\
{\it $^1$ Department of Physics, Tohoku University, 980--77 Sendai, Japan}\\
{\it $^2$ Faculty of Science and Engineering,}\\
{\it Ishinomaki Senshu University, 986 Ishinomaki,Japan}}
\date{}

\begin{document}
\baselineskip=9mm
\maketitle

\begin{center}
{\bf Abstract}
\end{center}

We calculate the electron screening effect
in low energy nuclear fusion reactions by taking
non-adiabaticity of the tunnleing process into account.
In order to investigate deviations from the adiabatic limit,
we use the dynamical norm method,
which has recently been developed by the present authors.
Using $d$+D reaction as an example, we show that the
screening energy never exceeds those estimated in the adiabatic approximation.
Our calculations indicate that
the non-adiabatic effect is important both in classically
allowed and classically forbidden processes.

\bigskip

\noindent
PACS number(s):
25.70.Jj, 
34.20.Cf, 
97.10.Cv, 
25.45.-z  

\newpage

Nuclear reactions at very low energies are important
for several problems of nuclear astrophysics.
An interesting problem is whether they
are significantly affected by bound or free electrons
when the reactions take place in laboratory
experiments or in condensed matters.
The screening effects by bound electrons in laboratory
experiments have been reported indeed
in ref.[1--8]. 
In these experiments, the target is in the
form of a neutral atom or molecule.
The electron clouds surrounding the target nucleus
screen the Coulomb repulsive potential between the colliding nuclei.
Consequently, the Coulomb barrier is reduced.
This leads to an enhancement of the fusion cross section and
the corresponding astrophysical $S$ factor.
This effect becomes significant at very low bombarding energies.
At such energies, the penetrability through the Coulomb
barrier exponentialy decreases with decreasing bombarding energy.
The barrier penetrability is, therefore, very sensitive to an even
tiny change of the potential barrier of the order of electron volt
due to the screening effects of the bound electrons.

Shoppa et al. studied the shift of the Coulomb barrier
due to the electron screening by solving the wave functions for electrons
in the classically allowed region.
In contrast to static estimates of the screening effects in
either the adiabatic or sudden approximations,
they investigated the incident energy dependence of the
screening effect\cite{SKLS93}.
Their calculations clearly show the transition from the adiabatic to the
sudden limits with increasing bombarding energy, and suggest the importance
of non-adiabatic effects at intermediate energies between the two
extreme limits.

However, they do not explicitly handle the tunneling process.
They assumed a constant energy shift inside the tunneling
region once they estimated it at the classical turning point.
In this paper we explicitly treat the
tunneling process and investigate the non-adiabatic effects
in the electron screening.
To this end, we use the dynamical norm method\cite{THA95}, which can be
applied to wide range of problems of quantum tunneling in systems with many
degrees of freedom lying between the adiabatic
and sudden tunneling limits \cite{CL81}.
This method first uses the tunneling probability in the adiabatic
limit as the reference. The effects of deviation from
that limit is then taken into account
through the reduction of the norm of the environmental
space during the classically forbidden process.

Our interest is to calculate the tunneling rate of
the relative motion between the colliding nuclei
in the presence of electronic degrees of freedom.
In particularly we consider in this paper the reaction
between deuterons($d$) and deuterium atoms(D),
for which experimental data on the possible effects of the
electron screening have recently been reported\cite{GGJRZ95}.
Moreover, this system is easier to be handled because
it contains only one electron, though
the dynamical norm method can be easily extended
to study more complex systems such as the experimentally
well studied $d$+$^3$He system, where there exist two bound electrons.

In the problem of electron screening, one often represents the
enhancement of the cross section in terms of the so called
screening energy $U_e$ defined by
\begin{equation}
U_e=\frac{E}{\pi\eta(E)}\log f
\end{equation}
where $f$ is the enhancement factor of the cross section i.e.,
the ratio of the cross section to that estimated for the
bare Coulomb barrier.
The bombarding energy in the center of mass frame
and the Sommerfeld parameter are denoted by $E$ and $\eta(E)$, respectively.
Equation (1) corresponds to assuming that the electron clouds
provide a constant energy shift $U_e$ of
the Coulomb barrier\cite{ALR87}.
In the adiabatic tunneling limit,
the screening energy is given by the difference in atomic binding
energies between the compound nucleus and the entrance channel\cite{BFMMQ90}.
For $d$+D reactions, the electron occupies the equally weighted
linear combination of the lowest energy gerade and ungerade molecular
orbitals in the entrance channel\cite{TYHP95}.
Therefore, the screening energy in the adiabatic approximation
reads
\begin{equation}
U_{ad}=\frac{E}{\pi\eta(E)}\log\left[\frac{1}{2}\left\{\exp \left(\pi\eta(E)
\frac{\Delta E_g}{E}\right)+\exp \left(\pi\eta(E)
\frac{\Delta E_u}{E}\right)\right\}\right]
\end{equation}
where $\Delta E_g$=40.7 eV is the difference of the binding energy
of electron in the 1s orbitals of He$^+$ atom and of D,
whereas $\Delta E_u$=0 eV
that in the 1p orbital of He$^+$ atom and
in the 1s orbital of D \cite{BFMMQ90}.
Note that the adiabatic screening energy $U_{ad}$
has the bombarding energy dependence.
This is a characteristic feature of the system with two identical
nuclei.

After discarding the center of mass motion of the whole system,
we choose the internuclear separation $\mbox{\boldmath $R$}$
and the electron center of mass location \mbox{\boldmath $r$},
which is measured from the center of mass of the two nuclei,
as two independent coordinates.
The Hamiltonian then reads
\begin{equation}
H=-\frac{\hbar ^2}{2\mu}\nabla^2_R
-\frac{\hbar ^2}{2m_e}\nabla^2_r
+\frac{e^2}{R}
-\frac{e^2}{\vert \mbox{\boldmath $r$}+\frac{\mbox{\boldmath $R$}}{2}\vert}
-\frac{e^2}{\vert \mbox{\boldmath $r$}-\frac{\mbox{\boldmath $R$}}{2}\vert}
\end{equation}
where $\mu$ and $m_e$ is the reduced mass between deuterons and the
electron mass, respectively.
The time dependent Schr\"odinger equation for the electron
in the external Coulomb field generated by two moving
deutrons is given by
\begin{equation}
i\hbar\frac{\partial}{\partial t}\psi(\mbox{\boldmath $r$},t)=
\left(-\frac{\hbar ^2}{2m_e}\nabla^2_r
-\frac{e^2}{\vert \mbox{\boldmath $r$}
+\frac{\mbox{\boldmath $R$}(t)}{2}\vert}
-\frac{e^2}{\vert \mbox{\boldmath $r$}
-\frac{\mbox{\boldmath $R$}(t)}{2}\vert}
+U_{ad}\right)\psi(\mbox{\boldmath $r$},t)
\end{equation}
Following ref. \cite{THA95}, we subtract the adiabatic
screening energy $U_{ad}$
from the potential energy of the relative motion between deuterons
and add it to the electronic Hamiltonian.
The initial wave function for the electron
$\psi(\mbox{\boldmath $r$},t=0)$
is the ground state of the deuterium atom boosted to the
correct center of mass velocity.
We assume that the electronic wave function is
azimuthally symmetric about the collision axis\cite{SKLS93}.
We use the method in ref.\cite{SB95} to perform the time
integration. This method modifies the Peaceman-Rachford method
\cite{KDMFKN77,KSK82} by incorporating with the time
expansion up to the second order of the
time step of the integration.

We solve eq.(4) from the initial position of $R$,
which we choose to be 10 a.u., to the outer classical turning point
by assuming that the relative distance between two deuterons
$R(t)$ obeys
\begin{equation}
\frac{\partial R}{\partial t}=-\sqrt{\frac{2}{\mu}
\left(E-(\frac{e^2}{R}-U_{ad})\right)}
\end{equation}
At the outer turning point, we switch the time to imaginary ($it\to\tau$)
and solve the time dependent Schr\"odinger equation along
the imaginary time axis
\begin{equation}
-\hbar\frac{\partial}{\partial\tau}\psi(\mbox{\boldmath $r$},\tau)=
\left(-\frac{\hbar ^2}{2m_e}\nabla^2_r
-\frac{e^2}{\vert \mbox{\boldmath $r$}
+\frac{\mbox{\boldmath $R$}(\tau)}{2}\vert}
-\frac{e^2}{\vert \mbox{\boldmath $r$}
-\frac{\mbox{\boldmath $R$}(\tau)}{2}\vert}
+U_{ad}\right)\psi(\mbox{\boldmath $r$},\tau)
\end{equation}
with
\begin{equation}
\frac{\partial R}{\partial\tau}=-\sqrt{\frac{2}{\mu}
\left(\frac{e^2}{R}-U_{ad}-E\right)}
\end{equation}
We solve these equations up to the inner turning point, which we
assume to be twice of the deuteron radius.
Note that eq.(6) describing the time evolution of the electronic
wave function does not conserve the norm of the wave function.
Following the dynamical norm method, the tunneling probability
for the inclusive process is given by $P(E)=P(E+U_{ad})\cdot\cal{N}$,
where $\cal{N}$ is the norm of the
electronic wave function at the inner turning point\cite{THA95}.
The enhancement factor is then given by
\begin{equation}
f=f_{ad}\cdot{\cal N}=\exp\left(\pi\eta(E)\frac{U_{ad}}{E}\right)
\cdot\cal{N} \\
\end{equation}
The dynamical norm factor $\cal{N}$ represents the deviation
from the adiabatic tunneling limit.

Figure 1 shows the enhancement factor of the barrier penetrability as a
function of the bombarding energy. The solid line is the result of the
dynamical norm method (see eq.(8)), while the dotted line is that
in the adiabatic approximation. Over the whole range of the bombarding
energy shown in this figure, the adiabatic approximation
overestimates the enhancement factor.
As we have remarked in refs.\cite{THA95,THAB94},
the adiabatic approximation gives the upper bound of the
tunneling rate.

Figure 2 shows the screening energy obtained from the enhancement
factor according to eq.(1).
The meaning of the solid and the dotted lines is the same as in fig.1.
The dashed line correspponds to the result reported in ref.\cite{SKLS93},
where the screening energy was estimated by studying the electronic
wave function only in the classically allowed region.
In all calculations, the screening energy decreases as the bombarding
energy increases.
The large difference between the dotted and the solid lines show that
the non-adiabatic effect is significant in the tunneling region.
Further, we notice that the dynamical norm method
(the solid line) gives a significantly smaller screening energy
than that in ref.\cite{SKLS93} (the dashed line).
This indicates that one needs to properly treat the tunneling region
in order to correctly estimate the screening energy.

In summary, we discussed the
non-adiabatic effect in the problem of the
electron screening in fusion reaction at low energies.
Comparison with the results in ref.\cite{SKLS93} shows that
the non-adiabatic effets are important both in classically allowed
and in classically forbidden processes.
We showed that the screening energy decreases with increasing
bombarding energy.
An unsolved puzzling problem in this field is that experimentally
observed screening effect is significantly larger than that
estimated in the adiabatic approximation\cite{L93,LL92}.
We showed that the non-adiabatic effects further reduce
the tunneling rate estimated in the adiabatic approximation.
Therefore the experimental enhancement of the fusion cross section
at extremely low energies in $d+D$ and other systems, where large
enhancement of the fusion cross section at low energies have been
reported like $d+^3He$,  might require additional ingredients
to the electron screening.
A possible candidate is the polarization of the colliding
deuteron\cite{B95}. This problem will be reported in a separate paper.

\bigskip

The authors thank J. Kasagi, Y. Kino, and A. Bonasera
for useful discussions.
The work of K.H. was supported by Research Fellowships
of the Japan Society for the Promotion of Science for
Young Scientists.
This work was supported by the Grant-in-Aid for General
Scientific Research,
Contract No.06640368, and the Grant-in-Aid for Scientific
Research on Priority Areas, Contract No.05243102,
from the Japanese Ministry of Education, Science and Culture.

\bigskip

\begin{center}
{\bf Figure Captions}
\end{center}

\noindent
{\bf FIG.1:} Enhancement factor of the barrier penetrability
as a function of the
bombarding energy for $d$+D reaction. The solid line is the
result of the dynamical norm method, which takes the reduction
of the tunneling rate due to the non-adiabatic effect into account.
The dotted line was obtained by using the adiabatic approximation.

\noindent
{\bf FIG.2:} Screening energy defined by eq. (1) as a function of the
bombarding energy. The meaning of the solid and the dotted lines is the
same as in fig.1. The dashed line was obtained by the same approach
as in ref.\cite{SKLS93}. The solid line explicitly treats
the dynamics in the tunneling region, while the dashed line
was obtained by studying the wave function of the electron
only in the classically allowed region.

\end{document}